\documentclass[aps,prd,showpacs,groupedaddress,nofootinbib]{revtex4}

\usepackage{color}
\usepackage{graphicx}
\usepackage{amsmath}
\begin{document}

\title{Reconstructing human organ cross-sectional imaging along any axis}

\author{Y. F. Fan$^{1,3}$\footnote{Corresponding authors\\ Email: tfyf@gipe.edu.cn}, Y. B. Fan$^{2,3}$, L. P. Luo$^{4}$, W. T. Lin$^{1}$, Z. Y. Li$^{5}$, X. Zhong$^{4}$, C. Z. Shi$^{4}$, T. Newman$^{5}$, Y. Zhou$^{1}$, C. S. Lv$^{1}$ and Y. Z. Fan$^{5}$}
\affiliation{
$^{1}$Center for Scientific Research, Guangzhou
Institute of Physical Education, Guangzhou 510500, P.R. China\\
$^{2}$Bioengineering Department, Beijing University of
Aeronautics and Astronautics, Beijing 100191, P.R. China\\
$^{3}$Key Laboratory of Optimal Design and Evaluation of Medical Equipment and Rehabilitation Aids, BUAA Research Institute, Guangzhou 510530, P.R. China\\
$^{4}$Medical Imaging Center, the First Affiliated Hospital of Jinan University, Guangzhou 510632, P.R. China\\
$^{5}$College of Foreign Studies, Jinan University, Guangzhou 510632, P.R. China}

\date{\today}

\begin{abstract}
Cross-sectional imaging of human organ serves as a critical tool to provide diagnostic results of many diseases. Based on a unique body coordinate system, we present a method that we use to reconstruct any cross-sectional imaging of organ regardless of its original section going along which scanning or cutting axis. In clinical medicine, this method enables a patient to undergo only one scanning, and then the doctor can observe the structure of lesion sections along any axis, and it can help find changes of lesions at the same section from different scanning results and thus quantify diagnosis by cross-sectional imaging. Significant progress has thus been made towards quantitative diagnosis cross-sectional imaging.
\end{abstract}


\maketitle

\section{Introduction}
In this paper, cross-sectional imaging (CSI) refers to magnetic resonance imaging and computed tomography (CT) imaging of human organ as well as sliced imaging of such. Be it in vivo or in vitro, CSI is playing an increasingly important role in medicine. Clinically, it is still considered to be the gold standard in diagnosing many diseases \cite{1,2,3}. In research area, CSI analysis is ubiquitous \cite{4,5,6,7}. A large number of facts \cite{8,9,10,11,12} show that CSI plays a key role in describing and predicting organ morphological changes.

However, a major problem with CSI as diagnosing standard is that when we want to observe the process of tissue/organ disease changes, how we can make sure that the position and posture of the organ from two scans to be the same; or after the tissue is sliced, how we can possibly observe the CSI of an organ going along the cutting axis. It is not advisable to resolve this problem directly by changing the scanning or cutting methods because we cannot make the subject keep exactly the same posture for two scans, nor can we restore an organ that has been cut.

The CSIs can be used to create a virtual organ by stacking \cite{13,14,15,16,17}. A virtual organ is composed of a finite number of volume elements. The morphology and structure of a virtual organ can be expressed by the geometric relationship between volume elements. Cutting a virtual organ requires positioning, so a body coordinate system needs to be established.

We assume that a virtual organ has a body coordinate system that is irrespective of how the scanning or cutting is taken, and that any section along any axis can be reconstructed with this unique body coordinate system, and when being reconstructed, its original morphology and structure can be kept. In this study, we intend to apply the principal axis of inertia of the asymmetrically shaped and anisotropically structured organ to develop its body coordinate system \cite{18}. Our experimental results show that the body coordinate system can be used to position an organ, irrespective of its scanning position or posture; when the body coordinate system is used as a reference system, the CSIs along any axis can be reconstructed, keeping its original morphology and structure. This method can be used to reconstruct the scanned organs in vivo, and the cut organ in vitro.

\section{Materials and methods}

\subsection{Equipment}
The scan equipment was Brilliance 64-slice Scanner by Philips, Netherlands, provided by Image Processing Center of Zhujiang Hospital. Scan settings were: frame bone tissue; power: $120 kv$; layer distance: $0.45 mm$, width: $768 pxl$, height: $768 pxl$. The scanning was conducted along both feet transect, from top to bottom. The subjects were asked to remain in $anatomical position$. A stereolithography 3D printer (Eden $260 V^{TM}$, Objet Eden $260 V$) and Objet VeroWhitePlus RGD835 were used.

\subsection{Participants}
The subjects were six healthy male wrestlers from the Guangdong (China) Provincial Sports School. Before the scan, each subject's medical history was reviewed and all the subjects were $x$-rayed to exclude subjects with diseases such as foot pathological changes, deformity or injury. Brain images were retrieved from http://bigbrain.cbrain.mcgill.ca/.

\subsection{Positioning of the virtual organ}
The position of the organ's center of mass (COM) is obtained by the following equation:
\begin{equation}
\label{eqn-1}
\left( x_{c}=\frac{\sum^{n}_{i=1} x_{i}\rho_{i}\Delta V}{\sum^{n}_{i=1} \rho_{i}\Delta V},y_{c}=\frac{\sum^{n}_{i=1} y_{i}\rho_{i}\Delta V}{\sum^{n}_{i=1}\rho_{i}\Delta V},z_{c}=\frac{\sum^{n}_{i=1} z_{i}\rho_{i}\Delta V}{\sum^{n}_{i=1}\rho_{i}\Delta V}\right),
\end{equation}
where $\left(x_{c},y_{c},z_{c}\right)$ stands for the organ's COM, $\left(x_{i},y_{i},z_{i}\right)$ stands for the volume element's position coordinate after the organ has been scanned, $\Delta V$ for the volume of volume element, $\rho$ for the density of volume element, and \emph{n} for the number of volume elements.

The rotating angle about axis \emph{x}:
\begin{equation}
\label{eqn-2}
\alpha=\frac{1}{2}\arctan\left( \frac{2\sum (y_{i}-y_{c})(z_{i}-z_{c})\rho_{i}\Delta V}{\sum (y_{i}-y_{c})^{2}\rho_{i}\Delta V-\sum (z_{i}-z_{c})^{2}\rho_{i}\Delta V}\right),
\end{equation}
where $\sum (y_{i}-y_{c})(z_{i}-z_{c})\rho_{i}\Delta V$ stands for the products of inertia rotating about axis \emph{x}, and $\sum (y_{i}-y_{c})^{2}\rho_{i}\Delta V, \sum (z_{i}-z_{c})^{2}\rho_{i}\Delta V$ for moments of inertia.

The positioning of rotation about axis \emph{x}:
\begin{equation}
\label{eqn-3}
\begin{pmatrix}
 x^{\alpha}_{i}\\
 y^{\alpha}_{i}\\
 z^{\alpha}_{i}\\
\end{pmatrix}=
\begin{pmatrix}
 1 & 0 & 0\\
 0 & cos(\alpha) & -sin(\alpha) \\
 0 & sin(\alpha) & cos(\alpha)\\
\end{pmatrix}
\begin{pmatrix}
 x_{i}\\
 y_{i}-y_{c}\\
 z_{i}-z_{c}\\
\end{pmatrix}+
\begin{pmatrix}
 0\\
 y_{c}\\
 z_{c}\\
\end{pmatrix},
\end{equation}
where $ x^{\alpha}_{i},y^{\alpha}_{i},z^{\alpha}_{i}$ sands for the volume element's position coordinate after rotating $\alpha$ about axis \emph{x}.

The rotating angle about axis \emph{y}:
\begin{equation}
\label{eqn-4}
\beta=\frac{1}{2}\arctan\left( \frac{2\sum (x^{\alpha}_{i}-x_{c})(z^{\alpha}_{i}-z_{c})\rho_{i}\Delta V}{\sum (x^{\alpha}_{i}-x_{c})^{2}\rho_{i}\Delta V-\sum (z^{\alpha}_{i}-z_{c})^{2}\rho_{i}\Delta V}\right),
\end{equation}
where $\sum (x^{\alpha}_{i}-x_{c})(z^{\alpha}_{i}-z_{c})\rho_{i}\Delta V$ stands for the products of inertia rotating about axis \emph{y}, and $\sum (x^{\alpha}_{i}-x_{c})^{2}\rho_{i}\Delta V, \sum (z^{\alpha}_{i}-z_{c})^{2}\rho_{i}\Delta V$ for moments of inertia.

The positioning of rotation about axis \emph{y}:
\begin{equation}
\label{eqn-5}
\begin{pmatrix}
 x^{\beta}_{i}\\
 y^{\beta}_{i}\\
 z^{\beta}_{i}\\
\end{pmatrix}=
\begin{pmatrix}
 cos(\beta) & 0 & sin(\beta)\\
 0 & 1& 0 \\
 -sin(\beta) & 0& cos(\beta)\\
\end{pmatrix}
\begin{pmatrix}
 x^{\alpha}_{i}-x_{c}\\
 y^{\alpha}_{i}\\
 z^{\alpha}_{i}-z_{c}\\
\end{pmatrix}+
\begin{pmatrix}
 x_{c}\\
 0\\
 z_{c}\\
\end{pmatrix},
\end{equation}
where $( x^{\alpha}_{i},y^{\alpha}_{i},z^{\alpha}_{i})$ has the same definition as that in Eq.~(\ref{eqn-4}), and $(x^{\beta}_{i},y^{\beta}_{i},z^{\beta}_{i})$ stands for the volume element's position coordinate after rotating $\beta$ about axis \emph{y}.

The rotating angle about axis \emph{z}:
\begin{equation}
\label{eqn-6}
\gamma=\frac{1}{2}\arctan\left( \frac{2\sum (x^{\beta}_{i}-x_{c})(y^{\beta}_{i}-y_{c})\rho_{i}\Delta V}{\sum (x^{\beta}_{i}-x_{c})^{2}\rho_{i}\Delta V-\sum (y^{\beta}_{i}-y_{c})^{2}\rho_{i}\Delta V}\right),
\end{equation}
where $\sum (x^{\beta}_{i}-x_{c})(y^{\beta}_{i}-y_{c})\rho_{i}\Delta V$ stands for the products of inertia rotating about axis \emph{y}, and $\sum (x^{\beta}_{i}-x_{c})^{2}\rho_{i}\Delta V, \sum (y^{\beta}_{i}-y_{c})^{2}\rho_{i}\Delta V$ for moments of inertia.

The positioning of rotation about axis \emph{z}:
\begin{equation}
\label{eqn-7}
\begin{pmatrix}
 x^{\gamma}_{i}\\
 y^{\gamma}_{i}\\
 z^{\gamma}_{i}\\
\end{pmatrix}=
\begin{pmatrix}
 cos(\gamma) & -sin(\gamma) & 0\\
 sin(\gamma) & cos(\gamma)& 0 \\
 0 & 0& 1\\
\end{pmatrix}
\begin{pmatrix}
 x^{\beta}_{i}-x_{c}\\
 y^{\beta}_{i}-y_{c}\\
 z^{\beta}_{i}\\
\end{pmatrix}+
\begin{pmatrix}
 x_{c}\\
 y_{c}\\
 0\\
\end{pmatrix},
\end{equation}
where $(x^{\beta}_{i},y^{\beta}_{i},z^{\beta}_{i})$ has the same definition as that in Eq.~(\ref{eqn-5}), and $(x^{\gamma}_{i},y^{\gamma}_{i},z^{\gamma}_{i})$ stands for the volume elements' position coordinate after rotating $\gamma$ about axis \emph{z}.

After being processed by Eq.~(\ref{eqn-2})-(7), the results will be processed again by Eq.~(\ref{eqn-2})-(7). When and ONLY when $\alpha=0,\beta=0,\gamma=0$, the operation will be completed. Since the principal axis of inertia has no direction while the coordinate axis has, the organ's posture varies after it has been positioned. We take the $anatomical posture$ as a standard to modify it. For instance, when the anteroposterior position of the brain is reversed after being positioned, we make it flip canvas horizontal for $180$ degree about the vertical axis.

\subsection{Reconstructing a virtual organ}
The reconstructed CSI is isotropic:
\begin{equation}
\label{eqn-8}\left\{
\begin{array}{c}
x^{'}_{i}=Round \left (\left ( \frac{\left (x_{i}-x_{c}\right)}{R},0\right)R \right)\\
y^{'}_{i}=Round \left (\left ( \frac{\left (y_{i}-y_{c}\right)}{R},0\right)R \right)\\
z^{'}_{i}=Round \left (\left ( \frac{\left (z_{i}-z_{c}\right)}{R},0\right)R \right)\\
\end{array}\right.,
\end{equation}
where \emph{R} stands for CSI resolution $(0.5mm)$, $(x^{'}_{i},y^{'}_{i},z^{'}_{i})$ for the coordinate of the reconstructed $(x_{i},y_{i},z_{i})$, and $(x_{i},y_{i},z_{i})$, $(x_{c},y_{c},z_{c})$ have the same definition as that in Eq.~(\ref{eqn-1}). The following is a CSI along one axis.

Along horizontal axis:
\begin{equation}
\label{eqn-9}\left\{
\begin{array}{c}
x^{'}_{i}=\left\{
 \begin{array}{ll}
 x_{i} & \hbox{$p+\Delta >x_{i}>p-\Delta$}\\
 space & \hbox{other}
 \end{array}
 \right.
\\
y^{'}_{i}=\left\{
 \begin{array}{ll}
 y_{i} & \hbox{other}\\
 space & \hbox{$x^{'}_{i}=space$}
 \end{array}
 \right.
\\
z^{'}_{i}=\left\{
 \begin{array}{ll}
 z_{i} & \hbox{other}\\
 space & \hbox{$x^{'}_{i}=space$}
 \end{array}
 \right.

\end{array}\right..
\end{equation}

Along frontal axis:
\begin{equation}
\label{eqn-10}\left\{
\begin{array}{c}
x^{'}_{i}=\left\{
 \begin{array}{ll}
 x_{i} & \hbox{other}\\
 space & \hbox{$y^{'}_{i}=space$}
 \end{array}
 \right.
\\
y^{'}_{i}=\left\{
 \begin{array}{ll}
 y_{i} & \hbox{$p+\Delta >y_{i}>p-\Delta$}\\
 space & \hbox{other}
 \end{array}
 \right.
\\
z^{'}_{i}=\left\{
 \begin{array}{ll}
 z_{i} & \hbox{other}\\
 space & \hbox{$y^{'}_{i}=space$}
 \end{array}
 \right.

\end{array}\right..
\end{equation}

Along vertical axis:
\begin{equation}
\label{eqn-11}\left\{
\begin{array}{c}
x^{'}_{i}=\left\{
 \begin{array}{ll}
 x_{i} & \hbox{other}\\
 space & \hbox{$z^{'}_{i}=space$}
 \end{array}
 \right.
\\
y^{'}_{i}=\left\{
 \begin{array}{ll}
 y_{i} & \hbox{other}\\
 space & \hbox{$z^{'}_{i}=space$}
 \end{array}
 \right.
\\
z^{'}_{i}=\left\{
 \begin{array}{ll}
 z_{i} & \hbox{$p+\Delta >z_{i}>p-\Delta$}\\
 space & \hbox{other}
 \end{array}
 \right.

\end{array}\right..
\end{equation}		

In Eq.~(\ref{eqn-9})-(11), \emph{p} stands for a point on axis, $p\in (0,\pm R,\pm 2R,\cdots)$, $(x^{'}_{i},y^{'}_{i},z^{'}_{i})$, $(x_{i},y_{i},z_{i})$  and \emph{R} have the same definition as those in Eq.~(\ref{eqn-8}), and $0<\Delta<\frac{1}{2}R$. This is to reconstruct a section.

\section{Results and Discussion}
\begin{figure}[!ht]
\begin{center}
\begin{tabular}{cccc}
 \includegraphics[width=12.8cm]{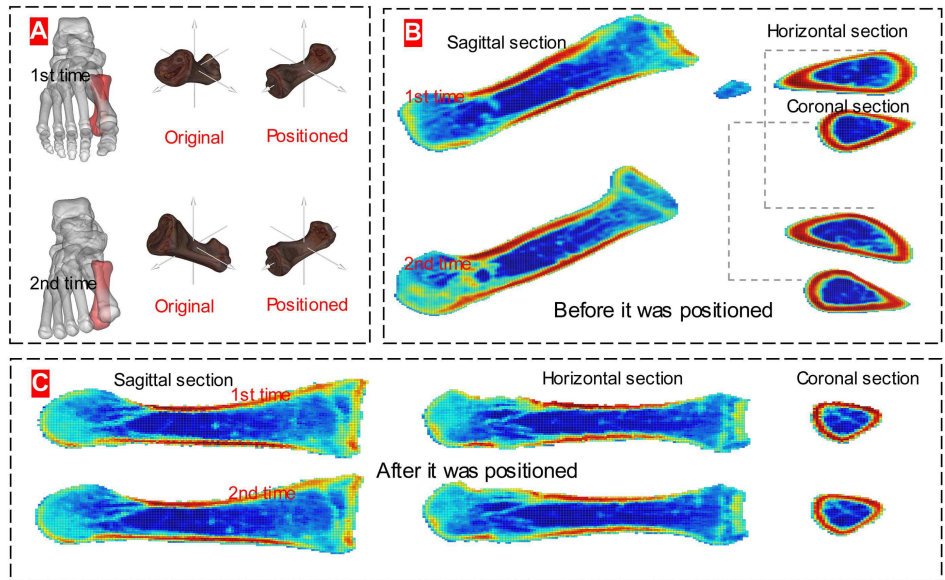}
\end{tabular}
\caption{\label{fig1} Standardized CSI of the first metatarsal of right foot. (A) Standardized body coordinate system of the first metatarsal. (B) COM on coronal, horizontal and sagittal planes before it was positioned. (C) COM on coronal, horizontal and sagittal planes after it was positioned. Eq.~(\ref{eqn-1}) is used to calculate the COM of the metatarsal, and Eq.~(\ref{eqn-9})-(11) to establish three sections of the scanned posture (with the COM of the metatarsal on the planes). See Fig.~\ref{fig1}A; Eq.~(\ref{eqn-1}) is used to calculate the COM of the metatarsal, Eq.~(\ref{eqn-2})-(7) are applied successively to position the metatarsal, Eq.~(\ref{eqn-8}) to reconstruct the CSIs (Fig.~\ref{fig1}B), and Eq.~(\ref{eqn-9})-(11) to establish three sections after they are positioned (with the COM of the metatarsal on the planes). See Fig.~\ref{fig1}C.}
\end{center}
\end{figure}

The standardization of an organ refers to the establishment of a coordinate system by using the organ's principal axis of inertia and by setting the organ's COM as the origin of the coordinate. The same subject's feet were scanned by 64-slice spiral CT scanner after an interval of $19 months$. The first metatarsal of his right foot was selected to be standardized. The results are shown in Fig.~\ref{fig1}.

Fig.~\ref{fig1}A and 1B indicate significant difference between the same subject's position and posture from two scans. Specifically, the first positioning was completed when the subject finished his first scan of the first metatarsal rotating about axis \emph{x}, \emph{y}, and \emph{z} (1st rotation  $(34.25,-10.31,22.48 3)$, 2nd rotation $(0.09,0.03,0.00)$); then the second positioning when rotating about axis \emph{x}, \emph{y}, and \emph{z} for the second time. For the second scan, the first positioning was made when the first metatarsal rotated about axis \emph{x}, \emph{y}, and \emph{z} (1st rotation $(38.34,14.92,5.86)$, 2nd rotation $(-0.04,0.00,0.00)$), and then the second positioning when the first metatarsal rotated again about axis \emph{x}, \emph{y}, and \emph{z}. The first metatarsal COM from the first scan was $(69.55,150.48,-111.94)$, and that from the second scan was $(70.30, 91.72, -168.04)$.

CSI analysis plays a vital role in clinical diagnosis because the CSI of an organ offers the most direct and reliable diagnostic evidence \cite{19}. But if we use the CSI directly from scanning, the results from Fig.~\ref{fig1}A question the reliability of such a diagnosis. The subjects take scans in anatomical posture, so when the interval between two scans is long, they might have forgotten their previous posture. In addition, other factors such as different operators, different facilities and the subjects' physical changes can all make it difficult to keep the same posture.

Fig.~\ref{fig1}C presents the reconstructed CSI, which is built upon the body coordinate system created by the principal axis of inertia of the virtual organ. For the same organ, the same CSI along the same axis from two scans can be reconstructed. The significance lies in the fact that clinically, doctors can accurately position the lesions and observe the changes of lesions at the same section from different scanning results so that they can give more precise and effective diagnosis. When the changes have been scanned for many times, the lesion changes can be clearly presented. After the CSIs from two scans are positioned, the doctor can confidently inform the patient that "comparing with the result from last time, we have a growth here, $\cdots$, which means $\cdots$". This method emerges as a reliable predictor in clinical diagnosis. When a photograph is taken under different inclination angles in crystallographic electron microscopy, 3D reconstruction of biological macromolecules can be built \cite{20}. The visualization of organ body coordinate system can bring about qualitative analysis of an organ's CSI diagnosis.

\begin{figure}[!ht]
\begin{center}
\begin{tabular}{cccc}
 \includegraphics[width=12.8cm]{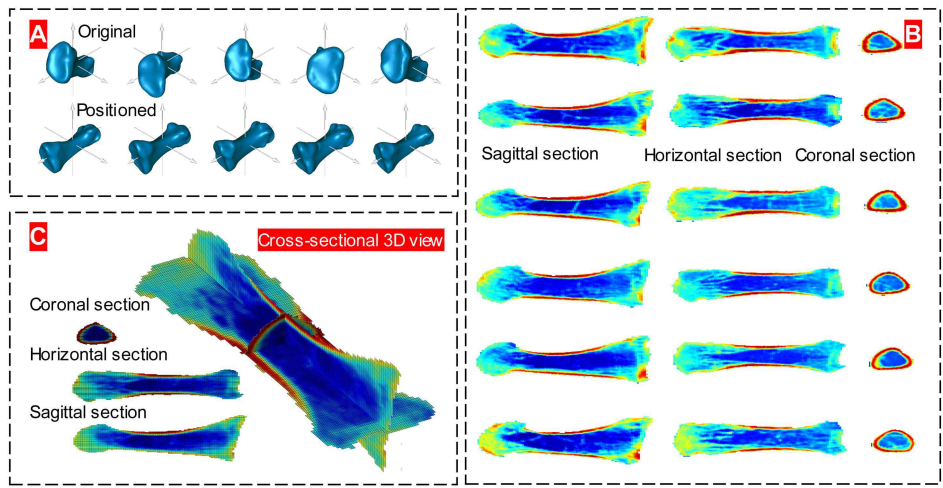}
\end{tabular}
\caption{\label{fig2} Merging of CSI of the first metatarsals of left foot. (A) Standardized body coordinate system of the first metatarsal of left foot. (B) Standardized geometry of the first metatarsal of left foot. (C) COM on coronal, horizontal and sagittal planes after the analogies were merged. The processing method is the same as that of the first metatarsal in Fig.~\ref{fig1}.}
\end{center}
\end{figure}

"Average face" can reveal face differences of people from different regions and areas. An "average organ" \cite{21} has its clinical significance because the morphology affects the structural function of an organ. A virtual organ is consisted of a finite number of cross sections, which means the modeling of an average organ should start from the average CSIs. In this study, six subjects' feet were scanned by 64-slice spiral CT scanner. The first metatarsals of their left foot were selected to be standardized. After being positioned, the CSIs were merged. The results are shown in Fig.~\ref{fig2}.

The subjects' first metatarsal basic morphological features and their position and posture information while being scanned are listed in Table~(\ref{t1}):
\begin{table}
\caption{\label{t1} Basic information on the first metatarsal}
\small
\begin{center}
\item[]\begin{tabular}{@{}*{8}{l}}
\hline
&No. &Volume$^{a}$ &Area$^{b}$ &Position$^{b}$ &1st rotation &2nd rotation\\
\hline
&1&12359.49&3674.19&(187.82,90.35,-162.21)&(38.80,-15.84,-3.13)&(-0.02,0.00,0.00)\\
&2&15135.75&4156.17&(194.25,105.35,-185.41)&(25.21,-25.58,-10.52)&(-0.15,0.03,0.00)\\
&3&12517.31&3732.70&(224.86,108.32,-215.00)&(-42.04,4.98,22.07)&(1.63,-30.84,-0.02)$^{*}$\\
&4&16644.79&4409.09&(183.93,101.91,-191.81)&(36.60,-37.49,0.07)&(-0.05,0.00,0.00)\\
&5&17006.04&4523.55&(208.98,107.08,-208.79)&(30.06,-23.61,-0.57)&(-0.01,0.00,0.00)\\
&6&18133.94&4591.07&(228.24,103.84,-200.65)&(43.94,-13.99,-15.57)&(-0.11,0.03,0.00)\\
\hline
\end{tabular}
\end{center}
\begin{quote}
$^{a}(mm^{3})$, $^{b}(mm^{2})$, $^{c}(mm)$. $^{*}$The 3rd rotation is (0.01, 0.00, 0.00).
\end{quote}
\end{table}

Fig.~\ref{fig2}B and Table~(\ref{t1}) show that the subjects' first metatarsal positions and postures are different from one another. Fig.~\ref{fig2}A and Table~(\ref{t1}) show that difference exists in each individual's organ morphology and the structure, indicating obvious personalization. This brings a clue to us: there is a great risk when comparing diagnoses made between CSIs of the same organ from each individual; for instance, when we compare CSIs from a healthy organ with those from a pathological one.

Clinically, however, the healthy people's many indexes are regarded as diagnostic standards. For example, temperature, body mass index, blood pressure, bone mineral density, and so on \cite{22}. This is also true to the CSIs. It is necessary to model an average CSI based upon differences between individuals and upon the requirements of a diagnosis. After the length and width vertical to the principal axis of a CSI have been rated by percentage, three average cross sections of the first metatarsal from six subjects are calculated (Fig.~\ref{fig2}C). It shows that the average cross section of the same organ from different subjects can be reconstructed. When the sample size is big enough, the average CSI of the virtual organ can bring clinical significance.

\begin{figure}[!ht]
\begin{center}
\begin{tabular}{cccc}
 \includegraphics[width=12.8cm]{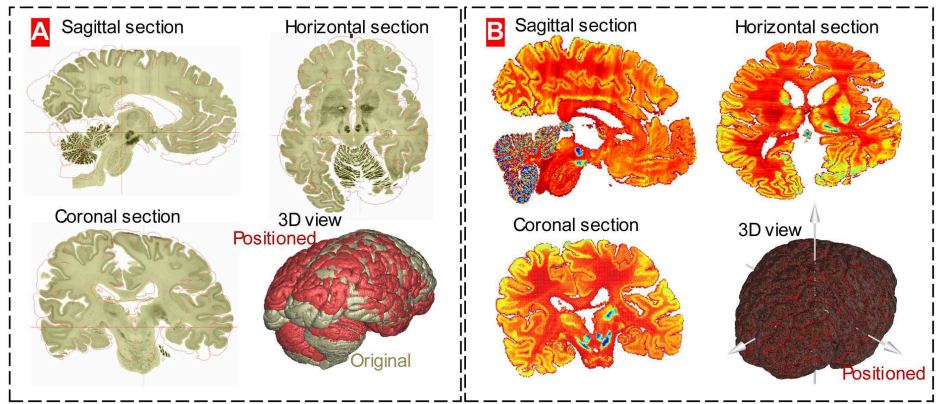}
\end{tabular}
\caption{\label{fig3} Standardized BigBrain. (A) COM on coronal, horizontal and sagittal planes before standardization. (B) COM on coronal, horizontal and sagittal planes after standardization. Change the image resolution from $6572*5711pixes$ \cite{17} to $328*285pixes$, and the slice distance from $0.02mm$ to $0.40mm$. Eq.~(\ref{eqn-1}) is used to calculate the COM of the BigBrain, Eq.~(\ref{eqn-9})-(11) to determine three sections of the scanned posture of the BigBrain COM (Fig.~\ref{fig3}A), Eq.~(\ref{eqn-1}) to calculate the COM of the BigBrain, and Eq.~(\ref{eqn-2})-(7) to position BigBrain, then Eq.~(\ref{eqn-8}) to reconstruct the CSIs, and finally Eq.~(\ref{eqn-9})-(11) to determine three sections after three BigBrain COMs are positioned (Fig.~\ref{fig3}B).}
\end{center}
\end{figure}

Theoretically, we may get the same position and posture of the same organ from two scans. In reality, however, we cannot observe the CSIs along the non-cutting axis after they have been cut. The downloadable original images from the BigBrain project \cite{17} offer us a chance to verify whether or not we can re-cut the reconstructed virtual organ that has been once cut. The virtual reality technology can be used to simulate an organ. Based upon the body coordinate system of BigBrain, the reconstructed CSI along non-cutting image is shown in Fig.~\ref{fig3}.

Fig.~\ref{fig3} shows that the virtual organ reconstructed by the CSI can again be reconstructed along any axis. The BRAIN (United States' Brain Research through Advancing Innovative Neurotechnologies) Initiative aims to create tools to image and control brain activity, while the European Union's Human Brain Project aims to create a working computational model of an organ \cite{23}. And these two programs will collaborate to understand how the brain functions, signaling the coming of a brain-research age. BigBrain is an important part of the European Union project. Thus, the sliced visible human and virtual organ have brought about a revolution to the modern medical education \cite{24,25} and they will prove to play an even more important role in the future brain age. If we want to observe the cross section along the non-cut axis (e.g. comparing one from the subject with that from the BigBrain), then we can reconstruct the CSIs along any axis.

The reconstruction of the CSIs along any axis can reduce the misdiagnosis rate. In the analysis of brain CSIs, Talairach and Tournoux introduced a coordinate system with the origin at the anterior commissure, a landmark, which has been identified in every brain \cite{26}. But one fact cannot be ignored, i.e. the effects from the position and posture of the brain when it is scanned. The origin at the anterior commissure is influenced by the position when being scanned, while the cross section will not only be influenced by the position, but also by the posture when being scanned (Fig.~\ref{fig1}A). Fig.~\ref{fig1}A and 2A show that whether it is the same subject or not, he/she cannot keep the same scanning posture for different times. This is true to the brain scan, especially when different subjects are being scanned from different facilities and from different hospitals. Comparative analysis is a common clinical application. In Boiselle's paper, when the same subject takes a scan three years later, the results (Fig.~\ref{fig1} and 2) \cite{27} suggest another risk of the CSI diagnosis, i.e. the misdiagnosis caused by the difference between two CSIs due to his scan postures. So we will analyze Fig.~\ref{fig1}C (the CSI of the same organ), not 1B.

Factual evidence is the only means to verify a scientific method \cite{28}. An object at hand is more persuasive. 3D printing enables us to do so. The following two examples of physical object demonstrate the reliability of this positioning method.
\begin{figure}[!ht]
\begin{center}
\begin{tabular}{cccc}
 \includegraphics[width=12.8cm]{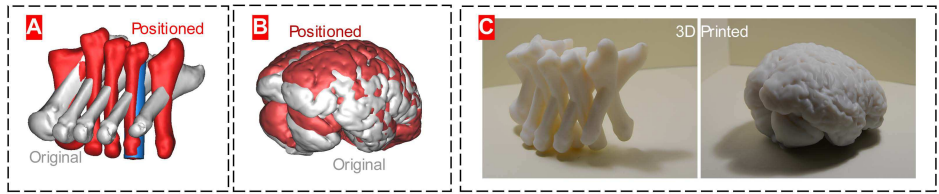}
\end{tabular}
\caption{\label{fig4} 3D printing of an organ. (A) The position and posture of left metatarsals in vivo before and after being positioned. (B) The position and posture of the BigBrain before and after being positioned. (C) 3D printing organs.}
\end{center}
\end{figure}

Fig.~\ref{fig4} verifies the reliability of our method. The deductive reasoning has proven the uniqueness of the body coordinate of the asymmetrically shaped and anisotropically distributed objects \cite{18}. Human organ is such an object \cite{29,30}. The key issue of this study is to examine whether the original morphology and structure of the organ will be kept the same after it is reconstructed based on its unique body coordinate. The physical objects from Fig.~\ref{fig4} are the results from 3D printing because only when an object's structure is continuum can it be printed.

Once the scanning axis of the equipment is fixed, the coordinate system of the scanning area is fixed. Then, the body coordinate system of an organ is determined by the position and posture of the subject when being scanned. There is not yet a method that can make the same subject keep the same posture in both scans. An ideal plan is to develop a method to construct a body coordinate system regardless of scanning posture or scanning or cutting axis. This study offers such an alternative. The supplementary animation file of the reconstruction of the CSI of the first metatarsal along any axis offers a more direct example.

The supplementary animation file shows the feasibility of the reconstruction of the CSIs of an organ going along any axis. After the organ is reconstructed upon the organ's unique body coordinate, its original morphology will be kept, so will be its structure. The CSIs along any axis just reflect the structure of an organ from different angles and different positions. The difference between our method and the visualization of the 3D reconstructed data set of the BigBrain from Atelier3D \cite{17} is that their body coordinate just translates on the organ, so we can only observe the cutting section and the other two sections vertical to the cutting section, while our method enables the body coordinate to rotate on the organ, so we can observe any cutting or scanning section along any axis because a virtual organ keeps its geometric invariance \cite{31}, such as rotation, or translation.

\section{Conclusion}
Choosing reference system and building coordinate system serve as the premises to describe the morphology of an organ and to predict its potential changes. For the Talairach and Tournoux coordinate system, measurements are taken from the CA-CP line, the vertical CA and the midline \cite{26}. The accuracy of their coordinate system relies on two factors - the accuracy of the brain's sagittal plane and the reliability of the landmarks. In our study, the organ's coordinate system is the result of calculation, i.e. a set of analytical solutions to the equations \cite{18}. This feature differs our method from the previous ones. Our method works for organs like brain as well as the first metatarsal. The other feature of this method is that when an organ's morphology and structure are kept the same, we may observe an organ's any CSIs along any axis.

The clinical significance of our research lies in the fact that with one scan result, the doctor can observe the morphology and structure of any CSI along any axis of the lesion, or the doctor can observe the morphological and structural changes at the same CSI of the lesion from a patient's different time-period scanning results. In this way, the misdiagnosis rate of the CSI will be reduced.

In addition, the possibility to re-cut the body coordinate might predict the birth of a real "virtual anatomy" \cite{25} to serve for medical education. When we instruct a human anatomy class, we do not have to observe the specimens soaked in formalin. Instead, we can take our own organ to understand its morphology and structure. And if the learning is blended with the organ's CSIs along any axis, it will be more interesting than just to observe the scanned CSIs or the sliced ones.

In summary, this paper has shown that the human virtual organ constructed upon CSI stacking will contribute to the positioning of human organ, which is irrespective of scanning or cutting axis, with the unique body coordinate of the organ established upon its principal axis of inertia, and to reconstructing the CSI along any axis upon the organ's body coordinate. However, attention should be paid to the limitation of the reconstruction of the CSI of a virtual organ, i.e. digitalization may lead to the insufficient image resolution. How to improve the reconstruction algorithm will be an important future research topic.

\section*{Acknowledgments}
This project was funded by the National Natural Science Foundation of China under the Grant Numbers $10925208$, $10972061$ and $11172073$. The authors would like to acknowledge the support from the subjects, Image Processing Center of Zhujiang Hospital, Image Processing Center of the First Affiliated Hospital of Jinan University.

\end{document}